\documentclass[preprint, superscriptaddress, showpacs, floatfix, nobalancelastpage]{revtex4}

\usepackage{amsfonts}
\usepackage{amssymb}
\usepackage[dvips]{graphicx}
\usepackage{color}
\usepackage{amsmath}
\usepackage[bookmarksnumbered, bookmarks, breaklinks, linktocpage]{hyperref}

\newcommand{\eend}      {\hspace{\stretch{1}}\rule{1ex}{1ex}}

\begin{document}

\title{Structural and magnetic instabilities of layered magnetic systems}

\author{Carsten~H. Aits}
\affiliation{Institut f\"ur Theoretische Physik, Universit\"at zu K\"oln, Z\"ulpicherstrasse 77,50937 K\"oln}
\author{Ute L\"ow}  
\affiliation{Institut f\"ur Theoretische Physik, Universit\"at zu K\"oln, Z\"ulpicherstrasse 77,50937 K\"oln}
\author{Andreas Kl\"umper}
\affiliation{Universit\"at Wuppertal, Gaussstr.20, 42097 Wuppertal, Germany}
\author{Werner Weber}
\affiliation{Institut f\"ur Physik,Universit\"at Dortmund,Otto Hahn-Str.4,D-44221 Dortmund, Germany} 

\date{\today}

\begin{abstract}
We present a study of the magnetic order and the structural stability 
of two-dimensional quantum spin systems in the presence of
spin-lattice coupling. For a square lattice it is shown that the 
plaquette formation is the most favourable form of static
two-dimensional dimerization.
We also demonstrate that such distortions may coexist with long range 
magnetic order, in contrast to the one-dimensional case. Similarly,
the coupling to Einstein phonons is found to reduce, but not to 
eliminate the staggered magnetic moment.

In addition, we consider the renormalization of the square lattice
phonon spectrum due to spin-phonon coupling in the adiabatic 
approximation. Towards low temperatures significant softening mainly 
of zone boundary phonons is found, especially around the $(\pi,0)$ 
point of the Brillouin zone. This result is compatible with the 
tendency to plaquette formation in the static limit.
We also point out the importance of a "magnetic pressure"
on the lattice due to spin-phonon coupling. At low temperatures, 
this results in a tendency towards shear instabilities of the lattice.

\end{abstract}

\maketitle

\section{Introduction}
Considering the immense wealth of materials  newly synthesized 
or found in nature  with low dimensional magnetic structure,
 one finds that the simple Heisenberg Hamiltonian,
 widely accepted as the paradigm of 
quantum magnets \cite{Heis,Bethe}, is often complicated by further interactions.
Many compounds thus show ground states and phases, different from the generic 
long range magnetic order of Heisenberg systems.

In this paper we study
the coupling of the magnetic degrees of freedom  to lattice vibrations
as a prominent example of an interaction
``beyond the Heisenberg-Hamiltonian''.
This coupling, often 
referred to as "spin-phonon" interaction is generic to {\sl all} 
magnetic materials and thus the problem 
to what extent the magnetic structure 
is changed by its presence concerns all 
low dimensional magnetic compounds.

Though spin phonon couplings are ubiquitous in nature, 
the possibilities of treating them  with 
some theoretical rigor are limited.
The main reason is that magnetic and phononic excitation energies are 
not well separated. 
In contrast, for electron phonon coupling we can dwell on the fact
that the bulk of the electronic excitations lies much higher
in energy than the phonon modes. This allows, in general, to use the 
adiabatic approximation when calculating the renormalization of 
phonon frequencies due to electron-phonon coupling.
For spin-phonon coupling, however, this concept cannot be applied,
in general.

One approach quite often used is to include local, Einstein-type 
phonons as quantum mechanical objects. This model may be applied for
certain vibrations of ligand atoms around the magnetic ions,
but it is not very realistic as it violates the infinitesimal 
translational invariance of the lattice.  

For relatively low-lying acoustic phonons, which are dominated by the
heavy magnetic ions, the response of the spin system may be
treated in adiabatic approximation.
In one dimension the paper of Cross and Fisher 
\cite{cross79} 
has treated the spin-phonon
coupling along such lines. In higher dimensions no study of spin-phonon
coupling in the adiabatic limit has so far been presented. 

In addition any effect of the magnetic pressure on the phonon vibrations
has been ignored so far.
The magnetic energy of a square lattice of spins with 1NN antiferromagnetic
coupling $J(a)$ is given as $E_m=U(T)J(a)$ which is typically
of the order $-0.1$ eV per atom. We may compare this magnetic energy
with the energy $E_{Mad}=-\alpha_M\frac{e^2}{a}$ 
of a cubic lattice of positive and negative elementary charges 
where the Madelung constant $\alpha_M$ is of order of unity. 
For rock salt we find e.g. $E_{Mad}\approx-8 eV$.
However, the pressure $\frac{\partial U}{\partial V}$ 
for the electrostatic system is 
$\approx 0.1 eV/ \text{\AA}^3$, while 
$\approx 0.015 eV /\text{\AA}^3$ 
for the magnetic system.
Although there is a factor of $\approx 100$ difference 
in the binding energies,
the magnetic pressure is only a factor $10$ smaller compared to the pressure of
the electrostatic system.
The main reason for the relatively large magnetic pressure
is the large spin phonon coupling caused by the strong distance
dependence of the super exchange $J(a)$,
which cannot be neglected in describing realistic quantum spin systems.

As an important special case we also discuss the 
limit of very strong spin phonon coupling, which leads to 
a dimerization of the lattice.
Typically, the response of the ground state energy of a two-dimensional 
magnetic system to dimerization is of second order and thus much weaker than 
in the one-dimensional case.
Two dimensional models with dimerization 
are thus essentially different compared to their one-dimensional analogues,
in particular in two dimensions dimerization does not
necessarily lead to a breakdown of magnetic long range order.
Also the  dimerization pattern, which leads to the lowest
ground state is not clear in two dimensions. 

This paper is organized as follows:
In Sect IIA we study statically dimerized models, in particular we
determine the optimal dimerization pattern in two dimensions. 
Sect IIB  is devoted to Einstein phonons coupled to two-dimensional
spin systems. 
In Sect III we calculate explicitly the phonon spectrum and study 
structural instabilities driven by the
spin-phonon coupling including the effect of 
the "magnetic pressure".

\section{Statically dimerized models and Einstein phonons}
\subsection{The dimerized model with optimal deformation patterns}
To approach the problem we 
assume in the first place 
that  in two dimensions similar to one dimensional systems 
the spin lattice interactions
lead to a transition to a dimerized state
i.e. to a pattern of strong and weak bonds
and we address the problem of the two dimensional correspondence
of a dimerized chain.

\begin{figure}[tp]
\centering  
\includegraphics[width=15.0cm,height=4cm]{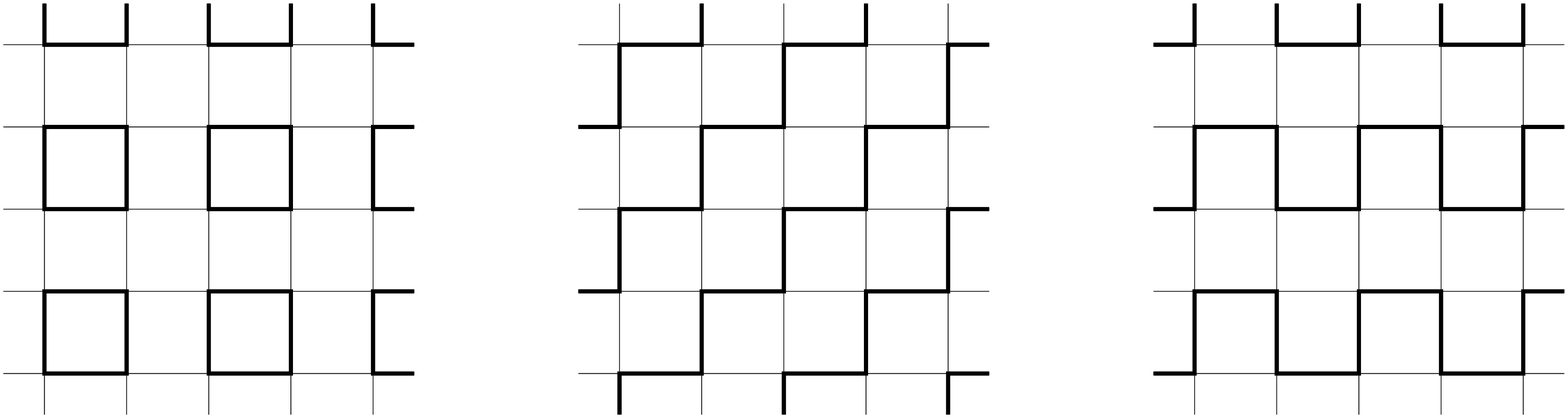}
\caption{Plaquette, stair, and meander configurations as 
possible dimerization patterns with minimal unit cell in two dimensions.} 
\label{fig1}
\end{figure}

Stair, plaquette and meander configurations (see Fig.\ref{fig1})
are obvious choices for such models,
which have been discussed controversially in the literature
\cite{tang88,KSH,Seng}. 

\begin{figure}[tp]
\centering  
\includegraphics[width=8.5cm,height=8cm]{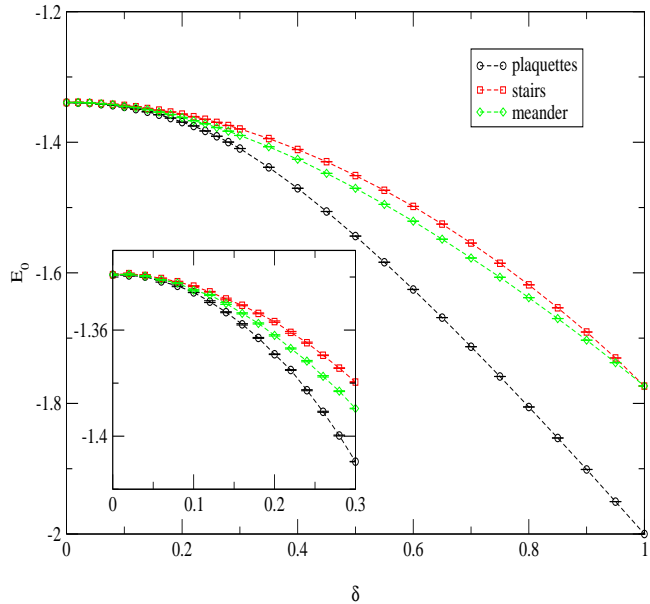}
\caption{Extrapolated ground state energies for the three dimerization
  patterns shown in Fig.\ref{fig1}.} 
\label{fig2}
\end{figure}

Here we show by a straightforward Monte-Carlo calculation \cite{remark1}  that 
the plaquette-models have lowest energy. 
This is convincingly demonstrated in Fig.\ref{fig2}, where the extrapolated 
energies of the three configurations are shown.

It is not clear however, whether configurations with larger unit cells 
or even disordered systems can still have lower energies.
To clarify this point, we analyze Hamiltonians of the type
\begin{eqnarray}
  \label{eq:twod_dim_H}
H(\delta)= 2J\sum_{ij}\left[(1+A_{ij}\delta)\vec{S}_{ij}\vec{S}_{i+1,j}
          + (1+B_{ij}\delta)\vec{S}_{ij}\vec{S}_{i,j+1}\right] 
\end{eqnarray}
for all possible $A_{ij} ,B_{ij}= \pm1$,
with equal number of strong and weak bonds
$ M=\sum_{i,j} (A_{ij}+B_{ij}) =0.$ 
Here $A_{ij}$ and $B_{ij}$ specify the pattern and $\delta$ gives the 
strength of the deformation. This Hamiltonian of course includes
the configurations of Fig.1 as special cases.

This more general study stems from an expansion of the free energy of the 
system Eq. \ref{eq:twod_dim_H} in $\delta$
\begin{eqnarray}
F(\delta,T)=F(0,T)+\frac{1}{2} a(T) \delta^2 +\frac{1}{24} b(T) \delta^4 + O(\delta^6).
  \label{eq:free_exp}
\end{eqnarray}
Writing down $a(T)$ explicitly
we observe that it
can be viewed as Hamiltonian of 
a two layer Ising model with 
Ising spins $A_{ij}$  and $B_{ij}$   
\begin{eqnarray}
\label{eq:Ising_H}
a(T)=\frac{\partial^2 F}{\partial\delta^2}\bigg\vert_{\delta=0}
=4J^2 \sum_{ijd_1d_2} 
  \{ K^{xx}(d_1,d_2)&A_{ij}A_{i+d_1,j+d_2}&\ \\\nonumber
+K^{yy}(d_1,d_2)&B_{ij}B_{i+d_1,j+d_2}&\\
+K^{xy}(d_1,d_2)&A_{ij}B_{i+d_1,j+d_2}&
+K^{yx}(d_1,d_2)B_{ij}A_{i+d_1,j+d_2}\}\nonumber
\end{eqnarray}
where the couplings of the Ising model are given 
by  dynamic dimer-correlations 
\begin{eqnarray}
K^{qr}(d_1,d_2)=-\int_0^\beta d\tau\langle D^q_{00}(0)D^r_{d_1
d_2}(\tau) \rangle
\label{eq:dimer_corr}
\end{eqnarray}
evaluated in the two-dimensional Heisenberg model, i.e. for $\delta=0$. 
Here $q$ and $r$ are either $x$ or $y$ corresponding to dimer
operators $D^x_{ij}=\vec{S}_{ij}\vec{S}_{i+1,j}$ or
$D^y_{ij}=\vec{S}_{ij}\vec{S}_{i,j+1}$. 
This means the 
quantum nature of the model is incorporated in the 
long ranged Ising couplings which depend on the Euclidean dynamical dimer
correlation functions of the
isotropic Heisenberg model.
This "perturbative approach" is somewhat 
restrictive but we are interested in the phenomenologically
relevant small dimerizations. 
Also, with
some more numerical effort quadruple  and higher correlations could be 
studied and for large dimerizations the problem becomes 
rather trivial, since it is 
reduced to one dimensional Heisenberg chains whose properties are well known.

The dimer correlations Eq. \ref{eq:dimer_corr} of the 
Heisenberg model (as well as the data shown in Fig.\ref{fig2})  were 
evaluated using 
a quantum Monte Carlo (QMC) loop algorithm \cite{evertz93,evertz01}, 
based on a path integral representation of the partition
function \cite{suzuki77}. Applying the loop algorithm, autocorrelation effects
play a minor role due to global spin updates,
and expectation values of diagonal and
off-diagonal operators are calculated efficiently within the framework of
improved estimators \cite{brower98,alvarez00}.
Further, by taking the continuous time version of the algorithm 
\cite{beard96}, finite size effects in Trotter direction are avoided. 

With the dimer correlations 
Eq.\ref{eq:dimer_corr} 
as coupling constants of the Ising model
we performed a standard classical Monte-Carlo simulation 
together with  some  cooling procedure \cite{ul93,dk04}.
This analysis gives clear evidence for 
crossing stripe patterns in the plane of the A and B spins
as ground state of the Ising model \cite{remark2}. This 
amounts to a plaquette structure as lowest pattern of the model 
Eq.\ref{eq:twod_dim_H}. 

In contrast to the one-dimensional case where $a(T)$
is divergent \cite{KRS} when T approaches zero, $a(T)$
stays finite in two dimensions, which a posteriori justifies our ansatz.

A crucial point which further distinguishes the one dimensional
case from the two-dimensional is 
the presence of {\sl both} 
long range magnetic order 
and a finite dimerization  
in the two-dimensional plaquette-model,
whereas in one dimension any finite $\delta$  leads to
a spin liquid ground state with a gapped excitation spectrum.

For static phonons we definitely observe a breakdown of long range magnetic 
order for sufficiently strong deformation $\delta$. The critical value 
$\delta_c$ takes rather large and in general different values for different 
patterns. For the stair configuration $\delta_c$ takes its maximal value 1, at 
which the system decouples into one-dimensional subsystems. For the plaquette 
configuration, $\delta_c$ is considerably smaller than 1 (in the cluster 
series expansion approach $\delta_c \approx 0.3$ \cite{Koga}), while 
we determined from our data $\delta_c = 0.291(3)$.

\subsection{The effect of Einstein phonons in two dimensions}

To study the impact of a spin-phonon coupling on the
magnetic long range order we next consider a 
two dimensional model coupled to Einstein phonons. 
Although the model  is not 
realistic in many respects, it has still attracted a lot of attention 
and for the physics of its one dimensional form 
a fairly clear picture has emerged by now.
In particular it is well known, 
that the quasi long-range order, leading to a logarithmic 
divergence  in the structure factor at $q=\pi$ is destroyed 
by a relatively small spin-phonon coupling.
This rises the problem to what extent the strong long range order of
2d systems is influenced in the presence of Einstein phonons.
To answer this, we calculate the expectation value of the 
staggered magnetization operator,   
\begin{eqnarray}
  \label{eq:m_stag}
  \vec M_{st}&=&  \sum_{x,y} {(-1)^{(x+y)} \vec S_{x,y}}
\end{eqnarray}
which is  a measure of N\'eel-order in the ground state.
We adopt the common procedure (see \cite{Barnes, Sandvik}) 
and compute the expectation value of $\vec M_{st}^2$
\begin{eqnarray}
  \label{eq:m_stag_sq}
  M^2:=\frac{1}{N^4}\langle 0 | \vec M_{\text{st}}^2 |0 \rangle,
\end{eqnarray}
for the full quantum-Hamiltonian of
spins $\vec{S}_{ij}$ at sites $ij$ coupled to Einstein phonons $b^{x}_{ij}$,
and $b^{y}_{ij}$ by a spin phonon coupling $g$
\begin{eqnarray}
  \label{eq:H_full}
H&=2J \sum_{i,j=1}^{N}\vec{S}_{ij}\vec{S}_{i+1,j}(1+g[b^x_{ij}+b_{ij}^{x
 \dagger}]) \\ \nonumber
 &+2J \sum_{i,j=1}^{N}\vec{S}_{ij}\vec{S}_{i,j+1}(1+g[b^y_{ij}+b_{ij}^{^y \dagger}])\\
 &+\omega\sum_{i,j=1}^{N}( b_{ij}^{x \dagger} b^x_{ij} + b_{ij}^{y \dagger} b^y_{ij}). \nonumber
\end{eqnarray}

Employing the Monte-Carlo method, which was 
developed in Ref.\cite{kuehne,aits}, 
we find, as expected,
Heisenberg like behaviour in the case of
small 
spin-phonon couplings $g$ and  large values of the phonon frequency 
$\omega$. 
It is remarkable however that
for parameters
for which in the one-dimensional model one finds clear evidence for a finite
correlation length at T=0, the staggered magnetization
is again nonzero, and its extrapolated value is only about $10 \% $ 
reduced  compared to the two-dimensional Heisenberg model. 
Also,  $ M^2$  displays a dominant $1/N$-
finite size behaviour  \cite{CHN,HN,Sandvik} derived from the 
nonlinear sigma model description of the Heisenberg model.  
\begin{figure}[tp]
\centering  
\includegraphics[width=11.0cm,height=8cm]{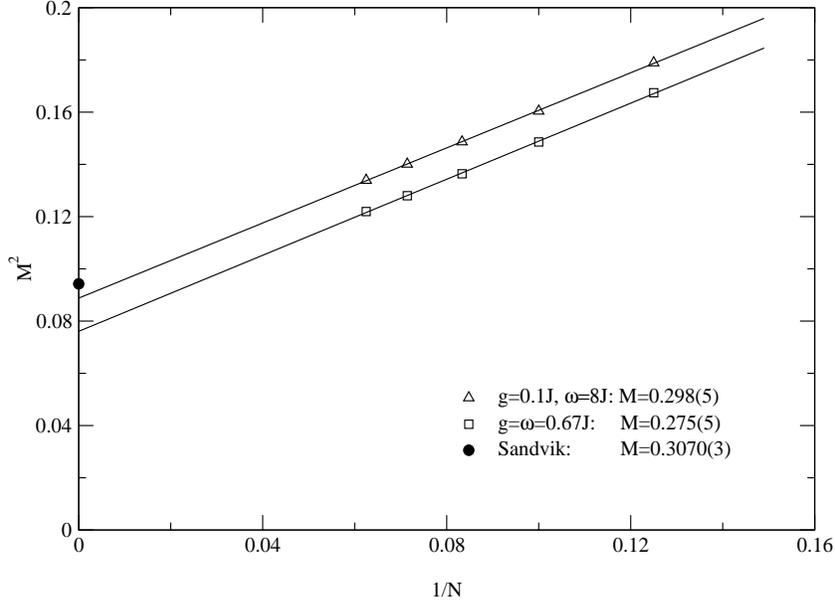}
\caption{Staggered magnetization $M^2$ for the model 
Eq.\ref{eq:H_full} as function of linear lattice size.
The black circle is the value for the Heisenberg-model
from Ref.\cite{Sandvik}.} 
\label{fig3}
\end{figure} 

Though we cannot exclude a breakdown of long range order 
by studying even larger spin-phonon 
couplings or different, maybe more elaborate coupling
mechanisms,  we find, that a realistic coupling strength 
comparable to the ones found in one dimensional magnetic materials
will {\sl not} lead to a breakdown of long range order.
So, the new feature we expect in two dimensions is the 
possibility of a structural phase transition
driven by spin-phonon interactions 
without breakdown of long range order.

\section{Structural phase transitions}
To investigate the structural phase transition also from 
a phenomenological point of view, 
we consider the following two-dimensional
Heisenberg-Hamiltonian coupled 
to classical phonons 

\begin{eqnarray}
H=2 J\sum_{l,m}&\frac{1}{2}\big[1+\lambda (\vec u_l-\vec u_m)\hat R_{lm}\big]\vec{S}_l \vec{S}_m,
 \label{eq:diff_coupling}
\end{eqnarray}

with $\vec S_l$ a spin $1/2$ operator at position $\vec r_l$,
$\vec R_{lm}= \vec R_l - \vec R_m$
the distance vector between sites $l$ and $m$ at equilibrium, 
$\hat R_{lm}= \vec R_{lm}/|\vec R_l - \vec R_m|$ and
$\vec u_l= \vec r_l + \vec R_l$
the displacement vector of the (l)th spin from equilibrium.
The summation in Eq.~\ref{eq:diff_coupling} runs over nearest 
neighbors l and m.  (Note that to simplify the notation
we abbreviate from now on the two spatial indices $(i,j)$ used 
e.g. in  Eq.~\ref{eq:twod_dim_H} by only one index and continue to use
the coupling strength $2J$ per bond.)

The total Hamiltonian consists of a sum of the Hamiltonian  
$H$ for all layers plus the phonon contributions $\Phi$,
which we assume to be derivable from a simple model with
central force potentials 
$\Phi_1(|\vec r_l -\vec r_m|)$ and $\Phi_2(|\vec r_l -\vec r_m|)$,
which depend on the nearest neighbor and the 
next nearest neighbor (diagonal) distances between the ions

\begin{eqnarray}
\Phi=\frac{1}{2}\sum_{l\neq m}\left(\Phi_1(|\vec r_l -\vec r_m|)+\Phi_2(|\vec r_l -\vec r_m|)\right).
  \label{eq:phonon}
\end{eqnarray}

Thus in equilibrium the total energy $E(a)$ per site is given by 

\begin{eqnarray}
E(a)=2 \Phi_1(a)+2 \Phi_2(\sqrt{2}a)+ 2 U(T) J(a)
\label{eq:energy}
\end{eqnarray}

where $U(T)<0$ is the internal energy per site of the Heisenberg model
with J=1 and  $a$ is the lattice constant.
The magnetic energy $U(T)J(a)$ has the tendency to compress the
lattice as the super exchange coupling $J(a)$, which may be derived from a
Hubbard-type model, increases with decreasing lattice constant a.
When we treat this magnetic pressure as an external pressure acting on the
lattice, the equilibrium condition $\frac{dE}{da}=0$ yields

\begin{eqnarray}
2B+4B^\prime + \frac{2}{a} J \lambda U(T) =0.
  \label{eq:equilib}
\end{eqnarray}

Here  $\lambda= \frac{d~ln J(a)}{d a}$ is again the 
spin phonon coupling, and following \cite{Mara} the force constants 
$B$ and $B^\prime$ are defined as

\begin{eqnarray}
B^{(i)}:= \frac{1}{R_{lm}}\frac{d \Phi^{(i)}}{dR_{lm}}
\end{eqnarray}

with $R_{lm}=|\vec R_l - \vec R_m|$ the equilibrium separation for
either first or second neighbor pairs. The full force constant matrix
$\Phi^{(i)}_{\alpha \beta}$ is given by \cite{Mara}

\begin{eqnarray}
\Phi^{(i)}_{\alpha \beta}  = A^{(i)} \hat R^\alpha_{lm} \hat R^\beta_{lm}
+B^{(i)}( \delta^{\alpha \beta}- \hat R^\alpha_{lm} \hat R^\beta_{lm})
\end{eqnarray}
with
\begin{eqnarray}
A^{(i)}=\frac{d^2}{dR^2_{lm}} \Phi^{(i)}(R_{lm}).
\end{eqnarray}

In any reasonable model of 1NN and 2NN force constants 
$A,B$ and $A^\prime, B^\prime$

\begin{eqnarray}
|B|<A \ \  \text{and}\ \  |B^\prime|<A^\prime
\end{eqnarray}
can be expected from the requirement that longitudinal phonon 
frequencies are in general considerably larger than transverse ones. 
One should keep in mind that our model for a square lattice
of magnetic ions ignores additional forces acting via the closed shell 
ligands, which may also contribute to the equilibrium condition 
Eq.\ref{eq:equilib}, so that without the magnetic pressure, $B+2B^\prime$ 
could be chosen positive. When, however, the magnetic pressure is
turned on, by lowering T, $B+2B^\prime$  will become negative (or
strongly reduced) which will affect, in particular, the stability of the
lattice against shear. 

\begin{figure}[tp]
\centering  
\includegraphics[width=11.0cm,height=8cm]{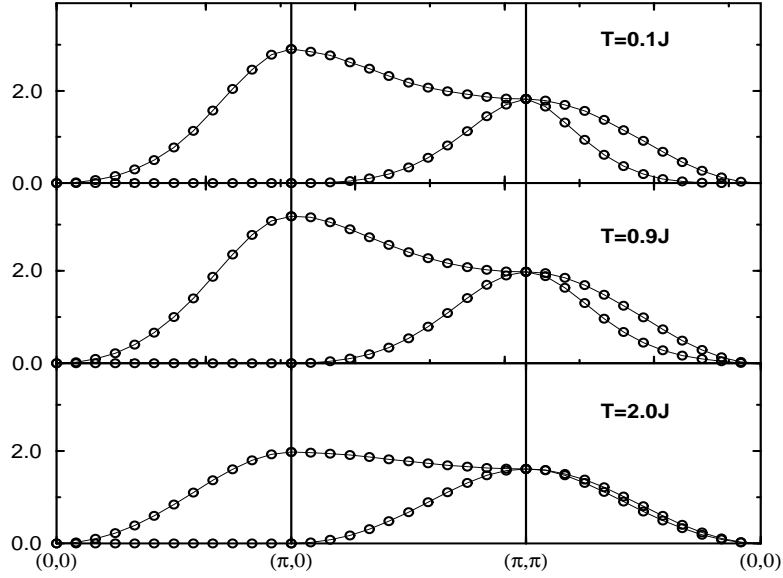}
\caption{Eigenspectra $\omega^2_{1,2}$ of the matrix $g$ for $T/J=0.1$,
$T/J=0.9$ and $T/J=2.0$
along a triangular path in $\vec{q}$-space.}
\label{fig4}
\end{figure} 

The elements of the dynamical matrix
$G_0^{\alpha \beta}(\vec{q})$ of the ``bare'' phonons are given by

\begin{eqnarray}
\label{first}
G_0^{xx}(\vec{q})&=& 2 A (1-\cos x)+ 2 B (1-\cos y) +2(A^\prime+B^\prime)(1-\cos x \cos y)\\\nonumber
G_0^{yy}(\vec{q})&=& 2 A (1-\cos y)+ 2 B (1-\cos x) +2(A^\prime+B^\prime)
(1-\cos x \cos y)\\
G_0^{xy}(\vec{q})&=& 2(A^\prime-B^\prime)\sin x \sin y\nonumber.
\end{eqnarray}

with $x=q_x a$ and $y=q_y a$.

We find the following phonon dispersion relations for longitudinal
and transverse modes.
For $\vec q=(x,0)$
\begin{eqnarray}
m& \omega_L^2(\vec q)=2( A+A^\prime +B^\prime)(1-\cos x)\\
m& \omega_T^2(\vec q)=2( B+A^\prime +B^\prime)(1-\cos x).
\end{eqnarray}

for $\vec q=(x,x)$
\begin{eqnarray}
m& \omega_L^2(\vec q)=2(A+B)(1-\cos x)+4A^\prime \sin^2 x\\
m& \omega_T^2(\vec q)=2(A+B)(1-\cos x)+4B^\prime \sin^2 x
\end{eqnarray}

At long wavelength, the transverse frequencies are given as
\begin{eqnarray}
m& \omega_T^2(x,0)=[(B+ 2B^\prime)+A^\prime - B^\prime)]x^2\\
m& \omega_T^2(x,x)=[2(B+ 2B^\prime)+A - B)]x^2.
\end{eqnarray}

In addition to the magnetic pressure, which is first order in $\lambda$,
the spin phonon coupling renormalizes the ``bare'' phonon frequencies similar
to the renormalization due to electron-phonon coupling.

Similar to the analysis described in the first part of
the paper, we again expand the free energy of the system Eq.\ref{eq:phonon}
in terms of the lattice displacement and 
find an expression, which is 
similar to Eq.\ref{eq:Ising_H}.
However here the $A_{ij}$ 
of Eq.\ref{eq:Ising_H}. 
are replaced by $u_{i+1,j}^x-u_{ij}^x$ and 
the $B_{ij}$ by $u_{i,j+1}^y-u_{ij}^y$.
That is compared to 
Eq. \ref{eq:Ising_H} the classical displacements 
are no longer Ising variables but take continuous values.
Thus to second order in the spin-phonon coupling we find for 
the dynamical matrix 

\begin{eqnarray}
G(\vec{q})= G_0(\vec{q})- \lambda^2 g(\vec{q}),
\label{eq:dyn_mat}
\end{eqnarray}
where the entries of the spin-phonon contribution 
\begin{eqnarray}
g^{xx}&= 2(1-\cos x)\widetilde{K}^{xx}(x,y),\\\nonumber
g^{yy}&= 2(1-\cos y)\widetilde{K}^{yy}(x,y),\\
g^{xy}&=g^{yx}=\frac{1}{2}(1-e^{ix})(1-e^{-iy})\widetilde{K}^{xy}(x,y)+ \text{c.c.}\nonumber
\end{eqnarray}

involve the Fourier transforms $\widetilde{K}^{qr}(x,y)$ 
of the dimer-dimer correlations Eq.\ref{eq:dimer_corr}.
Note that the matrix $g$ is real and symmetric, and from the symmetries of the 
dimer correlations  follows $\widetilde{K}^{xy}(\pi,\pi)=0$. 

Typical eigenvalues for the spin-phonon coupling matrix $g(x,y)$ along the
symmetry lines of the square lattice are shown in
Fig.\ref{fig4}. Several features are noteworthy:

\noindent
(i) there is a considerable T dependence not only concerning the magnitude
of the renormalization, but also the q-dependence.

\noindent
(ii) there is a maximum of the renormalization around $T/J=1$.

\noindent
(iii) along $(x,0)$ there is no renormalization of the transverse branch,
similar to the absence of the electron-phonon coupling for transverse modes in
a nearly free electron model.

\noindent
(iv) at small $T$, a significant splitting of the eigenvalues along
$(x,x)$ evolves, indicative of an increase of 
$\widetilde{K}^{xy}$ at low $T$.

\noindent
(v) also, the renormalization of the longitudinal branch 
along $(x,0)$ shows an increase of the effective $3NN$ force constant 
towards low $T$.

The points (iv) and (v) indicate that the effective forces between the atoms
transmitted via the spin system become longer ranged at low $T$.

\begin{figure}[tp]
\centering  
\includegraphics[width=11.0cm,height=8cm]{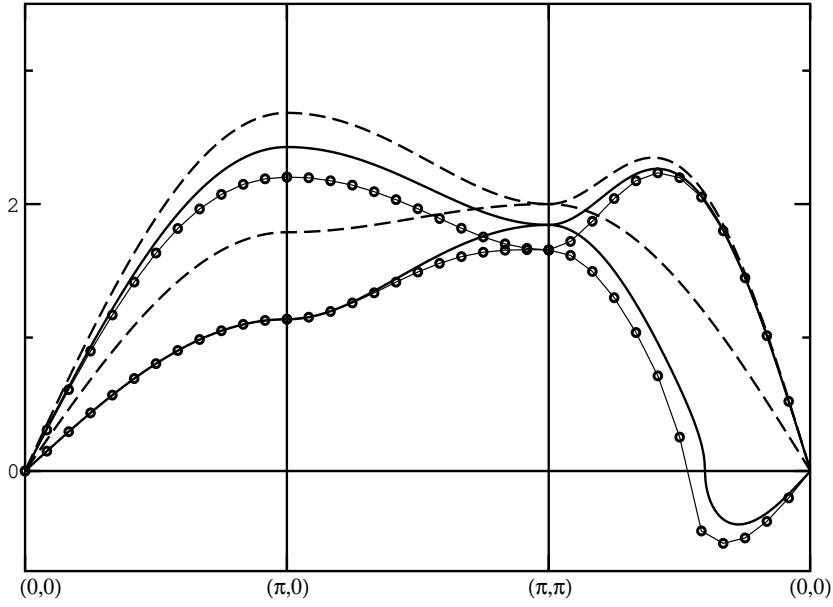}
\caption{Eigenspectra $\omega_{1,2}$  in zeroth 
(dashed line), first (solid line) and second order (circles)
along a triangular path in $\vec{q}$-space for 
spin-phonon coupling $\lambda=-0.6$,
$A^\prime /A=0.8$,  $B/J=-0.075$,  $B^\prime/J =-0.163$ and $T/J=0.1$.
Note, that in the region of instability with $\omega^2_{1,2}<0$
we plot $-\sqrt{-\omega_{1,2}}$.}
\label{fig5}
\end{figure} 

\begin{figure}[tp]
\centering  
\includegraphics[width=11.0cm,height=8cm]{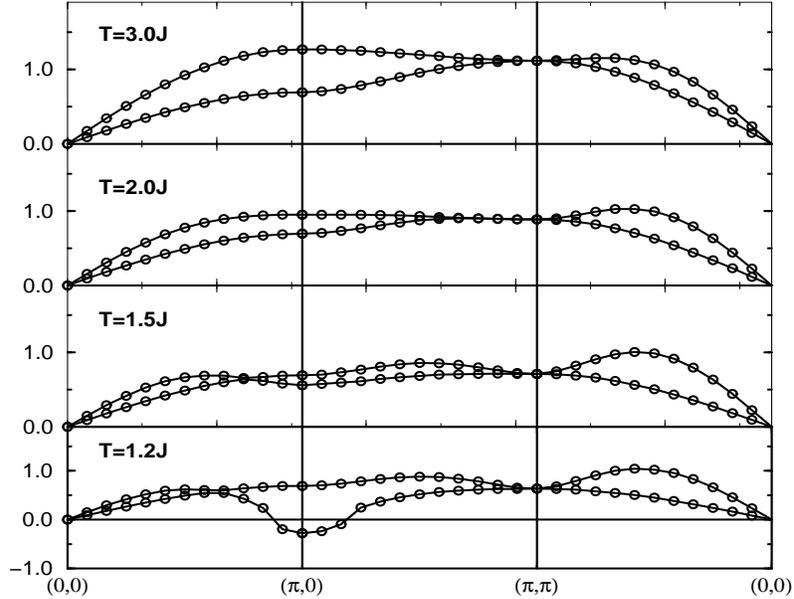}
\caption{Energies $\omega_{1,2}$ of the matrix $G$ for $B=B^\prime=0$. 
Again for $\omega^2_{1,2}<0$ we plot $-\sqrt{-\omega_{1,2}}$.}
\label{fig6}
\end{figure} 

The phonon renormalization due to spin-phonon coupling 
may lead to significant softening and even to lattice instabilities.

The magnetic pressure effect lowers specific transverse
phonon modes at long wavelengths, i.e. it destabilizes the lattice against 
certain shear deformations. This can be seen in  Fig.\ref{fig5}
where the undisturbed spectrum
of a square lattice (zeroth order in $\lambda$) which
follows from Eq.\ref{first}
by setting $ B=B^\prime=0$ is shown together  
with the spectra including the external pressure 
and the second order contribution.
Here the equilibrium condition Eq.\ref{eq:equilib}
is a strong constraint on the
system and the shear instability 
dominates the phonon renormalization.

We should keep in mind however that realistic planar quantum
antiferromagnets usually are ternary or even quaternary oxides and that 
the equilibrium condition of the lattice may be dominated by ligand
contributions, not included in Eq.\ref{eq:equilib}. As a consequence, we may
ignore the magnetic pressure effect and
consider solely the phonon renormalization due to the 
second order effect. Here primarily zone boundary phonons are
lowered, as may be seen from Fig.\ref{fig6}. For the calculation we have
 have chosen typical values for the model force constants, 
which approximately reproduce the acoustic phonon modes of 
typical cubic oxides with rock salt structure \cite{Kress} such as
MgO or BaO.

For a wide variety of $A^\prime /A$ ratios we find that the softening 
is strongest near $(\pi,0)$; only for somewhat pathological 
choices of $A, A^\prime, B, B^\prime$ such as  $A^\prime   -B^\prime \approx
0$, $A^\prime + B^\prime \gg A $,we find that the $(\pi,\pi)$ frequency goes
unstable first. The $(\pi,0)$ instability would lead to plaquette-like
distortions, while the $(\pi,\pi)$ instability would yield stair-case
distortions.

This means that our results from spin-phonon coupling agree with the
observation discussed above, that the plaquette is the  energetically most  
favorable dimerization pattern.

\section{Summary}

In this paper we have studied magnetic and structural instabilities
of two-dimensional quantum antiferromagnets. Of the various 
dimerization patterns 
the energetically most favourable one at $T=0$ is the plaquette order. 
It was also shown that the various dimerisation patterns do not necessarily 
lead to a breakdown of long range magnetic order.
In contrast to one dimension, the coupling to Einstein phonons
is found to reduce, but not to destroy the staggered 
magnetic moment.

 The tendency
to form dimerization patterns at sufficiently large spin phonon
coupling is also reflected in the large renormalization of specific
phonon frequencies, in particular near $(\pi,0)$ and $(\pi,\pi)$.
Our results indicate that again, plaquette formations are 
the most favourable structural distortions.

In this paper we also point out to the importance of a "magnetic
pressure" on the lattice caused by the relatively
large spin phonon coupling. This pressure effect builds up with 
decreasing $T$ and may lead to a significant decrease, if not instability of
specific shear elastic constants. Presently we have treated 
the magnetic pressure as an external one. In principle
it can also be included into the calculation of phonon frequency
renormalization, e.g. by carrying out so called frozen phonon
calculations, which is the ultimate form of the adiabatic approximation.

We finally remark that, wherever in the Brillouin zone there occur 
renormalization effects due to spin phonon 
coupling, also a significant increase of the phonon line width is expected.

\section{Acknowledgment}
We would like to thank D.Khomskii, E.M\"uller-Hartmann and J.Sirker
for useful discussions. This work was supported by the DFG through SFB 
608. Part of the simulations were carried out on the 
ZAM H\"ochstleistungsrechner J\"ulich.


\begin{thebibliography}{99}
\bibitem{remark1} Note that the error due to finite temperatures
is of the order of $T^3$ i.e smaller than $6.0\  10^{-5}$, we can therefore
within the statistical error safely identify U(T/J=0.04) with the ground state energy.
\bibitem{remark2} Note that the coupling of the Ising model depends on the
    temperature $T=1/\beta$ of the model Eq.\ref{eq:twod_dim_H}. We need to
    find the ground state of the Ising model for every T (though we are
    mostly interested in low T).The temperature used in the MC-simulation
    of the Ising model is an auxiliary variable.
\end{thebibliography}
\end{document}